# The mechanism of hole carrier generation and the nature of pseudogap- and 60K-phases in YBCO

K.V. Mitsen, O.M. Ivanenko

*Lebedev Physical Institute RAS, 119991 Moscow, Russia*

In the framework of the model assuming the formation of NUC on the pairs of Cu ions in $CuO_2$ plane the mechanism of hole carrier generation is considered and the interpretation of pseudogap and 60 K-phases in $YBa_2Cu_3O_{6+\delta}$ is offered. The calculated dependences of hole concentration in $YBa_2Cu_3O_{6+\delta}$ on doping $\delta$ and temperature are found to be in a perfect quantitative agreement with experimental data. As follows from the model the pseudogap has superconducting nature and arises at temperature $T^*>T_{c\infty}>T_c$ in small clusters uniting a number of NUC's due to large fluctuations of NUC occupation. (Here $T_{c\infty}$ and $T_c$ are the superconducting transition temperatures of infinite and finite clusters of NUC's, correspondingly). The calculated $T^*(\delta)$ and $T_c(\delta)$ dependences are in accordance with experiment. The area between $T^*(\delta)$ and $T_c(\delta)$ corresponds to the area of fluctuations where small clusters fluctuate between superconducting and normal states owing to fluctuations of NUC occupation. The results may serve as important arguments in favor of the proposed model of HTSC.

**The formation of NUC's in $YBa_2Cu_3O_{6+\delta}$**

Earlier [1-3] we proposed the mechanism of NUC formation in high-$T_c$ superconductors. The model is based on the assumption of rigid localization of doped charges in the close vicinity of doped ion. This localization results in local variation of electronic structure of the parent charge-transfer insulator that depends on the local mutual arrangement of the doped charges. It is shown that in such system NUC's may form under certain conditions on the pairs of neighboring Cu cations in $CuO_2$ plane.

Figure 1a shows the electronic spectrum of undoped high-$T_c$ superconductor. Here the charge-transfer gap $\Delta_{ct}$ corresponds to the transition of electron from oxygen to nearest Cu ion. The originating hole is extended over 4 surrounding oxygen ions (Fig. 1b) due to overlapping of their $2p_{x,y}$ orbitals. This formation ($3d^{10}$-electron+2p-hole) resembles a hydrogen atom. We have shown that the energy of two such excitations can be lowered (Fig. 1c) if such two side-by-side pseudo-atoms form a pseudo-molecule (Fig. 1d). This is possible due to formation of a bound state (of the Heitler-London type) of two electrons and two holes that emerge in the immediate vicinity of this pair of Cu ions. If we now decrease $\Delta_{ct}$ to the point where the gap disappears for two-particle transitions but remains for one-particle transitions, we arrive at a system in which some of the electrons belonging to the oxygen valence band effectively interact with pair states, or NUC's. We believe that the role of the doping is to activate possible NUC's.

As follows from [1-3] such NUC is formed in $YBa_2Cu_3O_{6+\delta}$ on a given pair of Cu ions in $CuO_2$- plane when three in a row oxygen sites in CuO-chain over (under) this Cu pair are occupied (Fig.2 a). The concentration of such triplets at random distribution of oxygen ions in chains is equal to $\delta^3$ per unit cell.

Isolated triplet of oxygen ions in chain forms two NUC's, one NUC in each of two $CuO_2$ planes (Fig.2a). If the number of consecutive oxygen ions in a chain $N_O > 3$, only every second triplet can form separated NUC (without common Cu ions) in each of $CuO_2$ planes (Fig.2 b). So, we may take that for $N_O > 3$ each triplet forms one NUC but only in one $CuO_2$ plane (Fig.2 b).

We will consider that a few of NUC's lying along a straight line in $CuO_2$ plane belong to 1D-cluster if all Cu ions included in NUC's form continuous cluster of sites. Accordingly, oxygen ions in chains forming a given 1D-cluster of NUC's form a continuous oxygen 1D-cluster in a plane of chains. So, for each 1D-cluster of NUC's in $CuO_2$ plane there is continuous generative cluster of oxygen ions in CuO-chain. The continuous sets of oxygen ions belonging to adjacent chains we assume to form united 2D-cluster of NUC's provided that they touch over 3 or more oxygen ions in adjacent chains (that is the percolation over NUC's takes place). This will correspond to the formation of continuous 2D-clusters of NUC's in both $CuO_2$ planes. Figure 3 shows the random distribution pattern for oxygen ions in chains for the 40×40 square lattice obtained by this method for $\delta=0.3$ and $\delta=0.6$.

The total number of NUC's in clusters (for both $CuO_2$ planes) per one unit cell of $YBa_2Cu_3O_{6+\delta}$ at random distribution of oxygen ions is equal to $N_U = \delta^3 + N_3(\delta)$, where $N_3(\delta)$ is the $\delta$-dependent number of isolated triplets of oxygen ions in chains equal to $N_3(\delta)=\delta^3(1-\delta)^2$. Respectively,

$$N_U(\delta)=\delta^3\{1+(1-\delta)^2\} \qquad (1)$$

It is seen that $N_U(\delta) \approx \delta^3$ for $\delta > 0.7$, when the bulk of NUC's belongs to large clusters.

At $\delta < \delta_c$ NUC's form finite clusters of various sizes. Within each cluster the NUC occupation number $\eta$ depends on temperature and equals [1,2] to

$$\eta = 2T/(T+T_0) \qquad (2).$$

As follows from (1) at $\delta > 0.7$ the volume concentration of NUC's $P = N_U/V_{UC} \approx \delta^3/V_{UC}$, where $V_{UC}=173\text{Å}^3$ is the volume of unit cell for $YBa_2Cu_3O_{6+\delta}$. Accordingly, the volume concentration of hole carriers $n$, generated in $CuO_2$ planes as electrons occupied NUC is equal to $n=\eta P=\eta\delta^3/V_{UC}=2\,(\delta^3/V_{UC})\,T/(T+T_0)$, and Hall coefficient is

$$R_H(\delta,T)=1\backslash ne=(1/2e)(V_{UC}/\delta^3)\,(T+T_0)/T, \qquad (3)$$

where $e$ is the electron charge. Figure 4a shows the temperature dependence of Hall coefficient for $YBa_2Cu_3O_{6.95}$ single crystals from [4], where authors first succeeded in separating of contributions from $CuO_2$ planes to Hall coefficient through the use of untwined $YBa_2Cu_3O_{6+\delta}$ single crystals with different $\delta$. As is seen from Fig. 4a the present data can be approximated successfully by (3) with $T_0\approx390$ K

Fig. 4b shows experimental $R_H(\delta)$ dependence for $T=300$ K from [35]. It is seen that experimental data are in a good agreement with the dependence obtained from (3) with $T_0\approx390$ K for $0.7<\delta<1$. In order to get the $R_H(\delta)$ curve for the whole range of $\delta$ the relation (1) for $N_U$ should be used. The experiment (Fig. 4 b) completely confirms this conclusion. It should be emphasized that the calculated curves on the Figures 4 a,b have no any fitting parameter except $T_0$, that just describes the temperature dependence $R_H(T)$ and allows to calculate $R_H$ values over the whole range of $\delta$ and $T$ variation.

The fact that hole concentration grows with a doping $\delta$ as $\delta^3$ can serve as powerful argument in favour of diatomic NUC existence in $YBa_2Cu_3O_{6+\delta}$ and as a support of the proposed mechanism of hole carrier generation in high-$T_c$ superconductors.

**Fluctuations and the nature of pseudogap. Phase diagram of $YBa_2Cu_3O_{6+\delta}$**

In [1,2] we have assumed that the pseudogap observed in different experiments is nothing but the same superconducting gap developing at $T>T_c$ due to the large fluctuations of NUC occupation because of electron transitions between the pair level of NUC and oxygen band.

The matter is in following. In conventional BCS superconductor with electron-phonon interaction the superconducting gap vanishes due to the thermal excitations over the Fermi surface, which decrease the number of unoccupied states available for the electron pair scattering. Analogously the mechanism of superconducting gap suppression in our case is the occupation of NUC's by real electrons. Therefore fluctuation-induced reduction of pair level occupation will amplify the superconducting interaction and can result in fluctuation-induced turning on superconductivity at $T^*>T>T_{c\infty}$ (here $T_{c\infty}$ is equilibrium value of $T_c$ for infinite cluster of NUC's). Opposite, the increasing of pair level occupation owing to fluctuation will reduce the superconducting interaction and can result in fluctuation-induced turning off superconductivity at $T_c<T<T_{c\infty}$. Large relative fluctuations of NUC occupation, corresponding to substantial deviation of $T^*$ and $T_c$ from $T_{c\infty}$ can happened in the underdoped samples when the significant number of NUC's belongs to finite nonpercolative clusters. The mean size of finite clusters decreases with doping reduction and relative fluctuations of NUC occupation increase in these clusters (i.e. $T^*$ goes up and $T_c$ goes down). In addition in overdoped samples when practically all Cu-ions belong to infinite percolation cluster, large fluctuations become impossible.

On the ground of the proposed model it is possible to deduce the dependences of $T^*$ and $T_c$ on doping $\delta$ for $YBa_2Cu_3O_{6+\delta}$. At $\delta<\delta_c$, when NUC's form finite clusters of various sizes, the sample should be defined as Josephson media, where superconductivity is realized over the whole volume thanks to the Josephson links between superconducting clusters.

The number of Cu ions included in cluster of NUC in $CuO_2$ plane we will take as a size of cluster, $S$. By the same time $S=3$ should be taken as a minimal cluster size where superconductivity could occur since it is impossible to consider the cluster with $S=2$ as superconducting due to the absence of superconducting transfer along cluster.

Let a cluster uniting some NUC's in $CuO_2$ plane contains $S \geq 3$ Cu-ions. According to (2), the number of electrons on NUC's in a given cluster at temperature $T$ is equal $N=TS/(T+T_0)$. This number can be changed on $\pm\sqrt{N} = \pm(TS/(T+T_0))^{1/2}$ due to fluctuations. The condition for turning superconductivity on (off) in a given cluster at temperature $T^*$ ($T_c$) is $N(T) \pm \sqrt{N(T)} = N_c$, where $N_c = T_{c\infty}S/(T_{c\infty}+T_0)$ – the number of electrons on NUC at the superconducting transition temperature $T_{c\infty}$ for the infinite cluster. Hence,

$$TS/(T+T_0) \pm (TS/(T+T_0))^{1/2} = T_{c\infty}S/(T_{c\infty}+T_0). \qquad (4)$$

Solving Eq.(4) with $T_0=390$ K and $T_{c\infty}=92$ K we find $T^*$ and $T_c$ as a function of $S$ (Fig. 5). As it seen from Fig. 5, the fluctuation effect on $T_c$ decreases with cluster size increasing and becomes to be negligible in clusters of NUC with more than 1500 Cu ions, that corresponds to the size ~150Å. The "plateau" at 60K on the curve $T_c(\delta)$, where $T_c$ changes between 50 and 70K in the range $0,6<\delta<0,8$, corresponds to S changing from 10 to 100. It is notable that there is minimal $S$ value, below that the cluster does not remain superconductive even at $T \to 0$ due to fluctuations of NUC occupation. Since the NUC occupation is $\eta \approx 2/5$ at $T = T_{c\infty}$, any fluctuation in cluster with $S<5$ that increases the number of electrons on NUC by 2 will result in the destruction of superconducting state. In order to find $T^*(\delta)$ and $T_c(\delta)$ dependences it is necessary to know $\delta_c$ corresponding to the percolation threshold over NUC and the statistics of finite clusters of NUC depending on $\delta$.

The percolation threshold over NUC as well as statistics of finite clusters can be found for the random distribution of oxygen atoms in chains by Monte Carlo method. According to the proposed mechanism of NUC formation we consider: 1) each 1D-cluster of oxygen ions in chain, containing $N_O \geq 4$ oxygen ions forms 1-D clusters of NUC with mean size $S = N_O - 1$ (i.e. containing $N_O - 1$ Cu-ions) in each of $CuO_2$ planes, 2) a size of 2D-cluster of NUC in $CuO_2$ plane is equal to the sum of sizes of constituent 1D-clusters. The value $\delta_c = 0,80 \pm 0,02$ has been determined by this method. It means that at $\delta > \delta_c$ $T_c$ would be equal $T_{c\infty}$. In the experiment however $T_c$ flattens out at $\delta > 0.85$ [5,6]. The increase of percolation threshold we suppose to connect with Cu-vacancies in chains and repulsion of oxygen atoms in adjacent chains [7], what prevents the growth of 1D-clusters. These factors will result in percolation threshold increasing in comparison with value expected for the random distribution of oxygen over the sites in chains.

To simplify the determination of $T^*(\delta)$ and $T_c(\delta)$ dependences we suppose that all finite clusters have the same sizes equal to some mean cluster size. The concept of mean cluster size $S_m$ is in use in the theory of percolation and is defined as weighted average $S_m = \sum n_i S_i^2 / \sum n_i S_i$. As it follows from the definition, large clusters give the basic contribution to $S_m$. And just this value $S_m(\delta)$ has to be substituted in (4) to find

$T_c(\delta)$ dependence because $T_c$ is determined as a superconducting transition temperature of large clusters with higher $T_c$, shunting small clusters and governing the basic contribution to conductivity and diamagnetic response.

Opposite, in order to get $T^*(\delta)$ the simple average $\bar{S} = \sum n_i S_i / \sum n_i$ has to be used since the contribution to fluctuation-induced turning on superconductivity is determined by the all finite clusters proportionally to their sizes. Figure 6 shows $S_m(\delta)$ and $\bar{S}(\delta)$ for 40x40 lattice obtained by Monte Carlo method. Evidently that $S_m$ tends to infinity as the percolation threshold approaches while $\bar{S}$ remains finite at $\delta \geq \delta_c$. Substituting the obtained values of $S_m(\delta)$ and $\bar{S}(\delta)$ in the quadric equation (4) we get $T_c(\delta)$ and $T^*(\delta)$ for $YBa_2Cu_3O_{6+\delta}$ as two solutions of this equation. Both solutions are shown in Fig. 7 by triangles up and down, correspondingly. Solid lines are drawn by eye. As follows from the model the area between these curves is the area of fluctuations, where the finite nonpercolative clusters fluctuates between superconducting and normal states due to fluctuations of occupations of NUC's. The dotted line of $T_c(\delta)$ at $\delta$<0.5 corresponds to the area where the mean size of cluster of NUC's $\bar{S}$ <5. As noted above, fluctuations will effectively destroy the superconductivity in these clusters. The experimental dependences $T^*(\delta)$ and $T_c(\delta)$ for $YBa_2Cu_3O_{6+\delta}$ single crystals are shown for comparison. Open squares are the data from [8], where the pseudogap opening temperature $T^*$ has been determined as the temperature of downward deviation of in-plane resistivity $\rho_{ab}(T)$ from the high-temperature $T$-linear dependence.. Open rhombuses corresponds to the temperature of superconducting transition $T_c$, determined by magnetic measurements in [6]. As is seen from the comparison of experimental dependences $T^*(\delta)$ and $T_c(\delta)$ with the calculated ones the agreement can be considered as good in spite of convention of their definition.

**Conclusion**

In the framework of the model assuming the formation of NUC on the pairs of Cu ions in $CuO_2$ plane the mechanism of hole carrier generation is considered and the interpretation of pseudogap and 60 K-phases in $YBa_2Cu_3O_{6+\delta}$ is offered. The calculated dependences of hole concentration in $YBa_2Cu_3O_{6+\delta}$ on doping $\delta$ and temperature are found to be in a perfect quantitative agreement with experimental data. As follows from the model the pseudogap has superconducting nature and arises at temperature $T^*>T_{c\infty}>T_c$ in small clusters uniting a number of NUC's due to large fluctuations of NUC occupation. The calculated $T^*(\delta)$ and $T_c(\delta)$ dependences are in accordance with experiment. The area between $T^*(\delta)$ and $T_c(\delta)$ corresponds to the area of fluctuations where small clusters fluctuate between superconducting and normal states owing to fluctuations of NUC occupation. The results may serve as an important arguments in favor of the proposed model of HTSC.

Correspondence and requests for materials should be addressed to K.V. Mitsen. (mitsen@sci.lebedev.ru).

Figure captions:

Fig. 1. (a) Electron spectrum of undoped cuprate high-$T_c$ superconductor. $U_H$ is the repulsive energy for two electrons on Cu ion. The charge transfer gap $\Delta_{ct}$ corresponds to the transition of electron from oxygen to nearest Cu ion with the origin of hole extended over four surrounding oxygen ions (b). (c) The energy of two such excitations can be lowered if two side-by-side "hydrogen" pseudo-atoms form "hydrogen" pseudo-molecule (d).

Fig.2. a) - the negative-U center (NUC) is formed in $YBa_2Cu_3O_{6+\delta}$ on a given pair of Cu ions in $CuO_2$ plane when three in a row oxygen sites in CuO-chain over (under) this Cu pair are occupied; b) formation of continuous NUC clusters in $CuO_2$ planes by the row of oxygen ions in CuO-chain.

Fig.3. Pattern of chain oxygen clusters forming a finite clusters of NUC's in $CuO_2$ planes obtained by Monte-Carlo method for random distribution of oxygen ions in chains: (a) -$\delta$=0.3 and (b) - $\delta$=0.6. Open circles are oxygen ions in chains. Clusters contained more than 3 oxygen ions are dashed.

Fig.4. Hall coefficient of $CuO_2$ plane of twin-free $YBa_2Cu_3O_{6+\delta}$ single crystal depending on temperature and doping: (a) open squares - $R_H(T)$ for $\delta$=0.95 [4]; (b) open rhombuses - $R_H(\delta)$ at $T$=300K [4]. Solid line on both figures is the dependence (4) with $T_0$=390 K. Thin line is the $R_H(\delta)$ dependence with regard to the additional contribution of isolated triplets of oxygen ions in chains.

Fig.5. $T^*$ and $T_c$ as a function of cluster size $S$ for 3<$S$<1500. Insert shows the same dependences for 3<$S$<100. The "60-K plateau" on $T_c(\delta)$ curve where $T_c$ changes from 50 K to 70 K corresponds to $S$ variation on order of $S$ magnitude (from 10 to 100).

Fig.6. The mean sizes of finite clusters of NUCs, $S_m$ and $\bar{S}$ versus $\delta$ for YBa$_2$Cu$_3$O$_{6+\delta}$. Open circles and squares are the results obtained by Monte Carlo method for 40×40 lattice for $S_m$ and $\bar{S}$ correspondingly. Solid lines are drawn by eye.

Fig.7. The comparison of calculated dependences of $T^*(\delta)$ (solid triangles down) and $T_c(\delta)$ (solid triangles up) for YBa$_2$Cu$_3$O$_{6+\delta}$. Open squares show the experimental results [8] for the single crystals, where $T^*$ has been determined as the temperature of downward deviation of in-plane resistivity $\rho_{ab}(T)$ from the high-temperature $T$-linear dependence. Open rhombuses are the results of the magnetic measurements of $T_c$ for YBa$_2$Cu$_3$O$_{6+\delta}$ single crystals [6]. Solid lines are drawn by eye. The dotted line of $T_c(\delta)$ at $\delta<0.5$ corresponds to the area where the mean size of cluster of NUC's $\bar{S}$ <5 and fluctuations effectively destroy the superconductivity in these clusters.

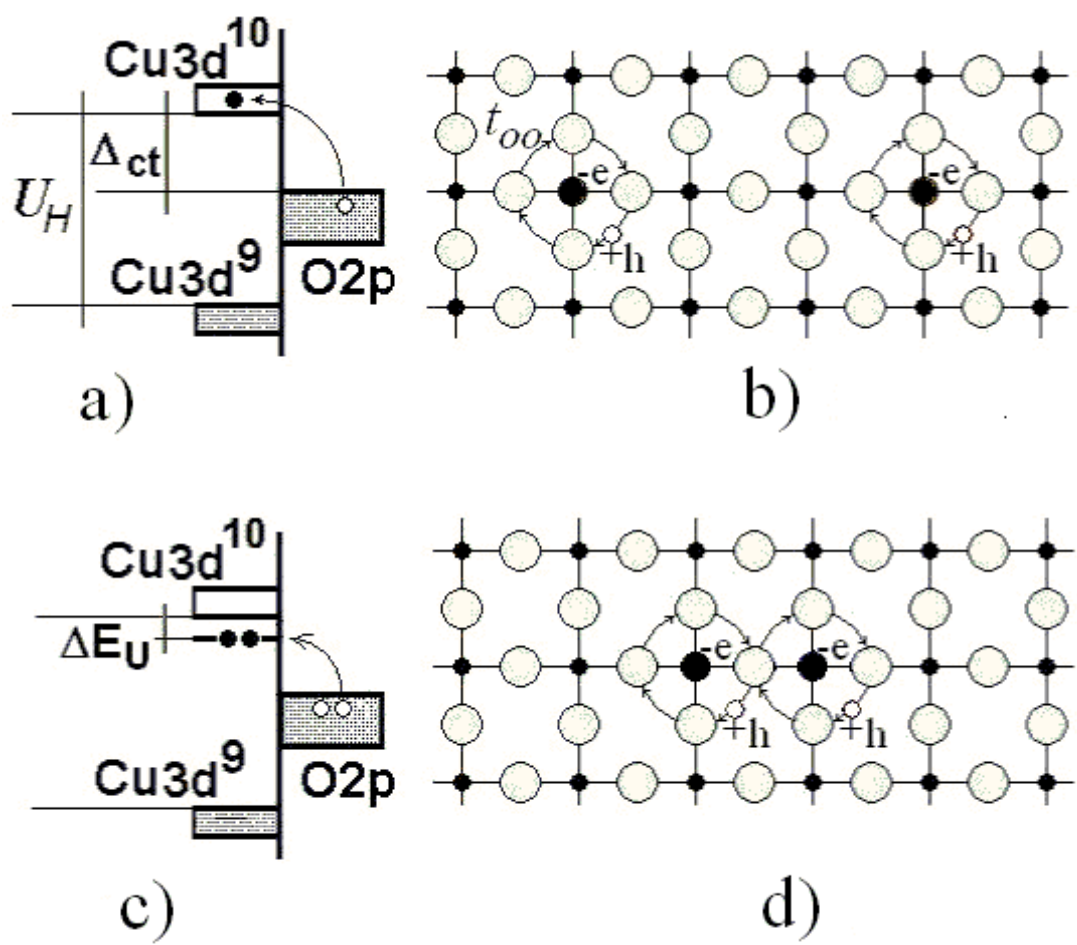

**Fig. 1**

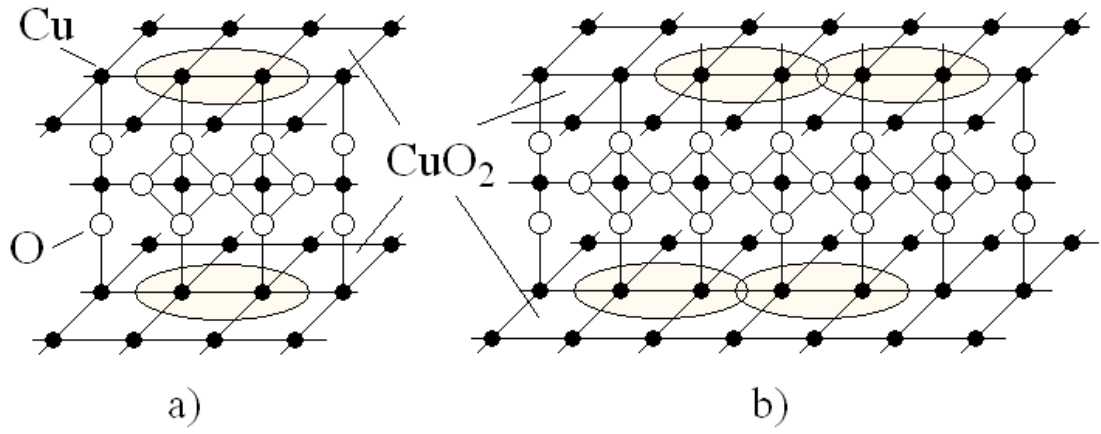

**Fig. 2**

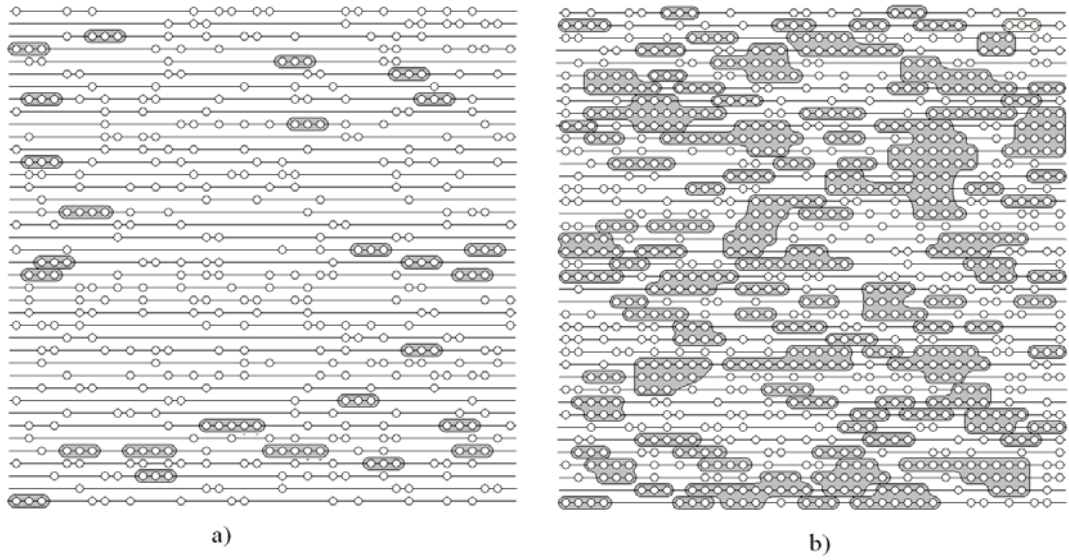

**Fig. 3**

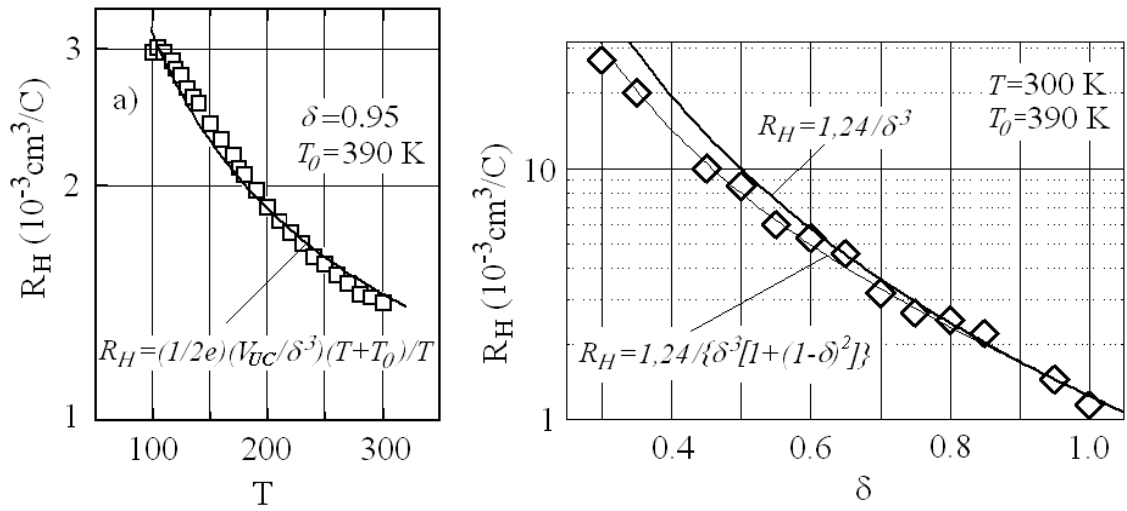

**Fig. 4**

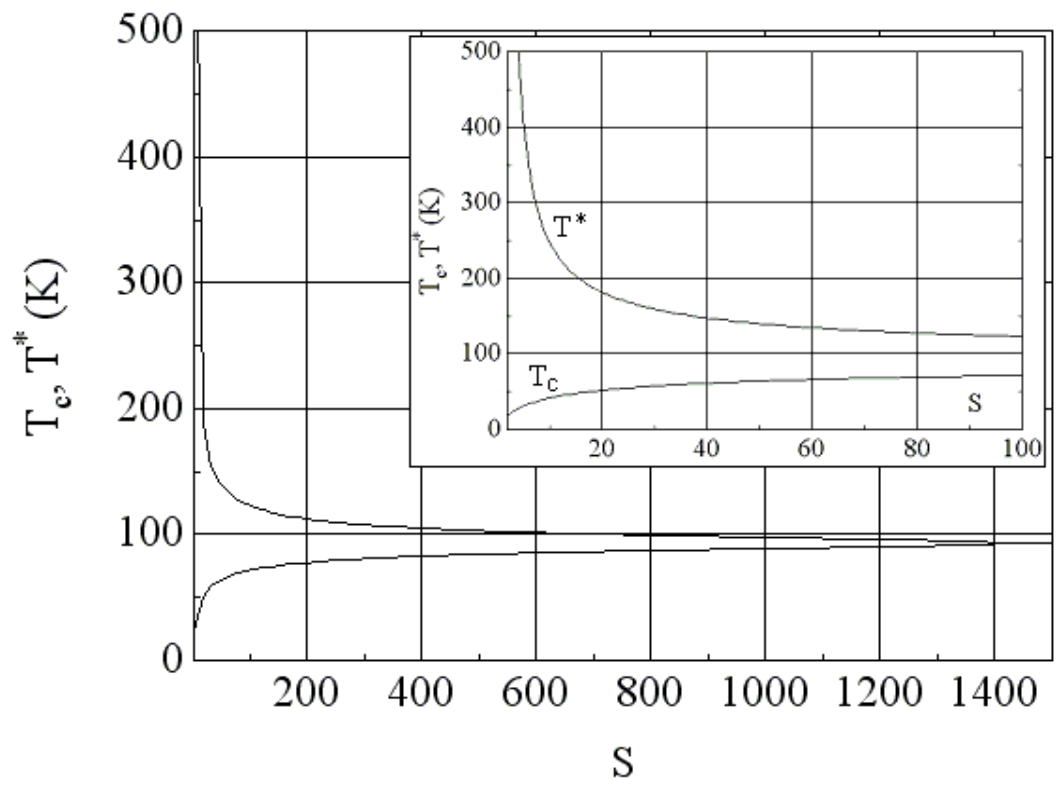

**Fig. 5**

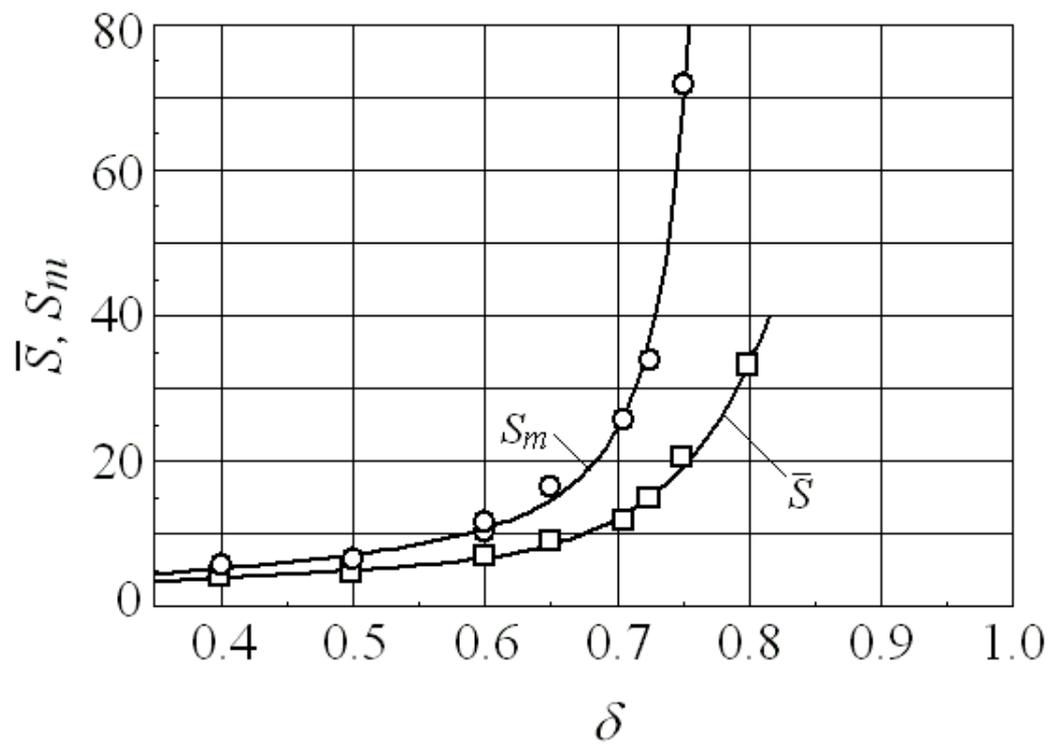

**Fig. 6**

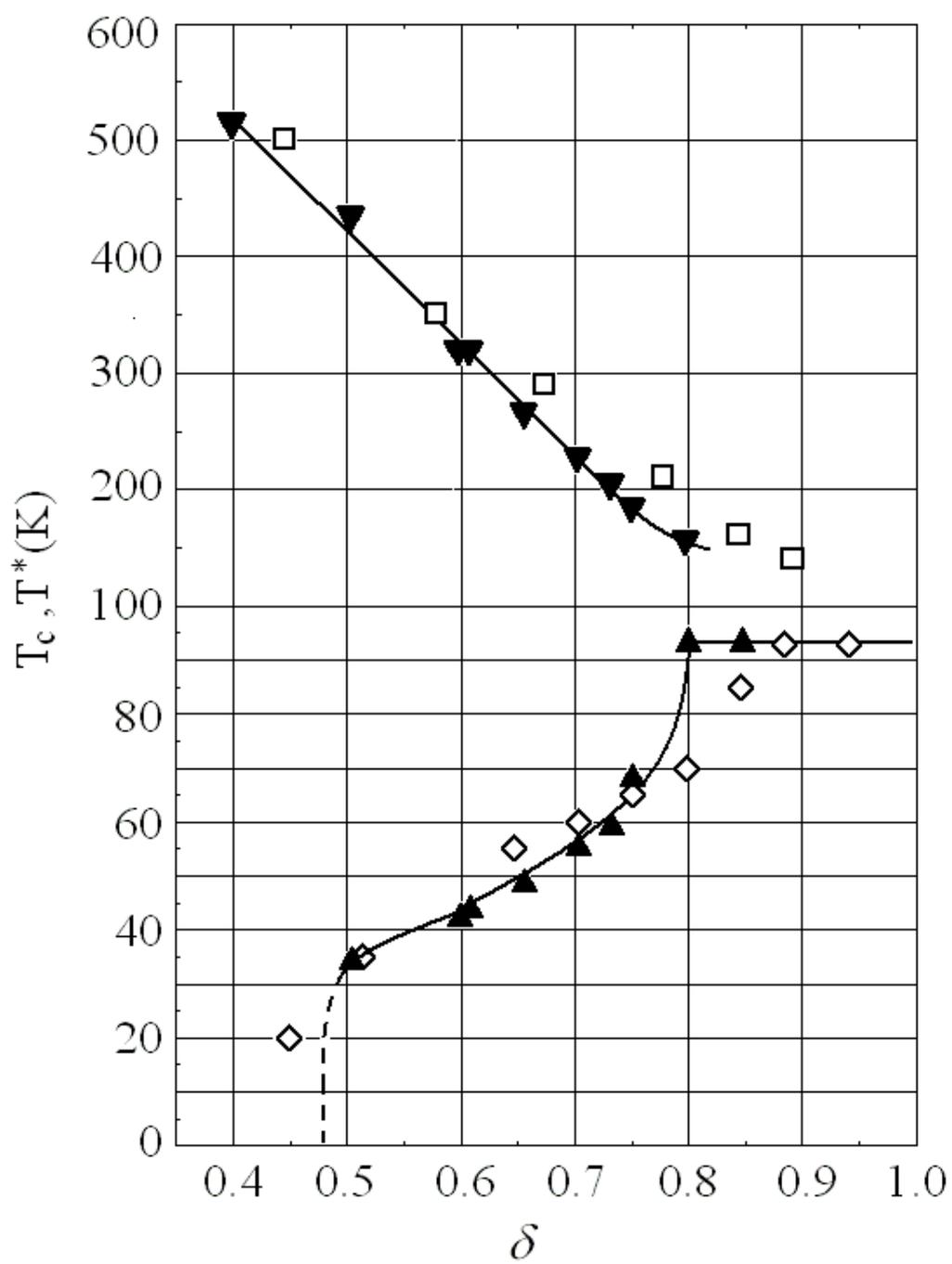

**Fig. 7**